\documentclass[12pt,thmsa]{article}
\usepackage{amsfonts}

\usepackage{sw20lart}

\input tcilatex
\begin{document}

\author{E. J. Doedel$\thanks{%
Supported in part by NSERC Canada Research Grant A4274},$ M. J. Friedman$%
\thanks{%
Supported in part by NSF DMS-9404912 and by NSF ATM 912-0321.}$ and B. I.
Kunin$\thanks{%
Supported in part by UAH Research Institute Mini-Grant}$ \\
$^{*}$Computer Science Department, Concordia University, Montreal,\\
Quebec, H3G 1M8 Canada \\
$^{\dagger ,\ddagger }$Department of Mathematical Sciences, University of
Alabama\\
in Huntsville, Huntsville, AL 35899}
\title{Successive Continuation for Locating Connecting Orbits. }
\date{}
\maketitle

\begin{abstract}
A successive continuation method for locating connecting orbits in
parametrized systems of autonomous ODEs is considered. A local convergence
analysis is presented and several illustrative numerical examples are given.
\end{abstract}


\textbf{AMS(MOS) subject classification:} \textit{34B15, 34C30, 34C35,
65J15, 65L10, 68-04.}

\section{Introduction}

The existence of a trajectory connecting equilibrium points of an ODE, a 
\textit{homoclinic orbit }or \textit{heteroclinic orbit}, also called 
\textit{connecting orbit is} of significance in a variety of applications.
Connecting orbits often arise as traveling wave solutions of parabolic and
hyperbolic PDEs., e.g., in combustions models \cite{bn}. They have been
shown to underlie intermittency phenomena in fluid mechanics \cite{ahls},
``bursting'' in models of biological cells \cite{re}, chaotic vibration of
structures \cite{m}, and chaotic behavior of electronic circuits, \cite
{cwhz, fra, frgp}, light pulses in fiber optics \cite{mt}, and chemical
reactions \cite{gs}. The corresponding numerical problem is that of finding
solutions $(u(t),\lambda )$ of the system of autonomous ODEs 
\begin{eqnarray}
&&u^{\prime }(t)-f(u(t),\lambda )=0,\;\;u(\cdot ),f(\cdot ,\cdot )\in \Bbb{R}%
^n,\lambda \in \Bbb{R}^{n_\lambda },  \tag{1.1a} \\
&&\lim_{t\rightarrow -\infty }u(t)=u_0,\;\;\lim_{t\rightarrow +\infty
}u(t)=u_1.  \tag{1.1b}
\end{eqnarray}
Most algorithms for computing connecting orbits reduce (1.1) to a boundary
value problem on a finite interval using linear or higher order
approximations of stable and unstable manifolds near $u_0$ and $u_1,$
respectively. We studied and applied such a method in \cite{df} and
generalized it in \cite{fd1}. An alternate convergence analysis of the
method is given by Schecter \cite{s}. The basic method of choosing
appropriate boundary conditions dates back to Lentini and Keller \cite{lk}.
A closely related procedure, which uses generalized eigenspaces for $%
f_u^T(u_0,\lambda )$ and $f_u^T(u_1,\lambda )$ to construct the relevant
projections, was developed by Beyn \cite{b1, b2} and used by Champneys and
Kuznetsov in their recent work on locating higher codimension homoclinic
orbit bifurcations \cite{ck}. A shooting method for computing connecting
orbits has been used by Rodr\'{\i}guez-Luis \textit{et al }\cite{rlfp}. For
parameter superconvergence results see the recent work of Sandstede \cite{sa}%
.

We considered the case of center manifold in \cite{f, fd2}. Champneys,
Kuznetsov, and Sandstede have also generalized the algorithm in \cite{ck} to
the case of nonhyperbolic equilibria \cite{cks}. Both, hyperbolic and
nonhyperbolic equilibria are considered by Canale \cite{c}, using multiple
shooting for discretizing the approximate truncated problem, together with a
new a new technique for evaluating partial derivatives in the boundary
conditions.

In this paper we present a local convergence analysis of the \textit{%
successive continuation method} for \textit{locating} connecting orbits.
This method aims at computing a connecting orbit when there is a region of
uncertainty for the parameter values, which is the case in most practical
problems. In contrast, most previous work is concerned with the \textit{%
continuation }of connecting orbits, given a sufficiently accurate starting
connecting orbit.

Assume that $u_0$ and $u_1$ are hyperbolic fixed points of the vector field $%
f$, and that the local unstable manifold $W_{loc}^u(u_0)$ at $u_0$ has
dimension $n_0$, and that the local stable manifold $W_{loc}^s(u_1)$ at $u_1$
has dimension $n_1.$ Let $S_0$ and $S_1$ denote the tangent planes to $%
W_{loc}^u(u_0)$ and $W_{loc}^s(u_1),$ respectively. In the finite interval
approximation, the scaled independent variable takes values in $[0,1].$ A
general outline of the full algorithm for locating and continuing connecting
orbits is then as follows:

(I) \textit{Time integration to get an initial orbit. }Typically, this gives
a very inaccurate ``approximation'' of a connecting orbit with either $%
(u(0)-u_0)\in S_0$ or $(u(1)-u_1)\in S_1$. This step can be done by
continuation with the ``period'', i.e., the total integration time, as
parameter.

(II) \textit{Locate a connecting orbit. }Use a sequence of homotopies
(``successive continuation'') that lead to an orbit with both $(u(0)-u_0)\in
S_0$ and $(u(1)-u_1)\in S_1$.

(III) \textit{Increase the accuracy.}\textbf{\ }Use continuation to further
increase the period of the approximate connecting orbit.

(IV) \textit{Continue the connecting orbit}.\ If desired\textit{, }compute
an entire family of approximate connecting orbits.

In this paper we are primarily concerned with Step~(II). In Section 2 we
formulate the approximate problem and a successive continuation algorithm.
This algorithm is implemented in a general code based on the continuation
package AUTO \cite{dwf} and is used to compute the examples in Section 4
below. In Section 3 we prove a local result, namely that the successive
continuation algorithm has an open neighborhood of convergence. More
precisely, we show that for any point in this neighborhood there exists a
continuous piecewise smooth homotopy leading to the solution. Section 4
contains examples. In Section 5 we conclude with a discussion of the
applicability and limitations of the algorithm.

In our original algorithm for continuing connecting orbits \cite{df, f} we
included the eigenvectors as unknowns in the full system to be solved at
each continuation step. This can lead to an ill-conditioned system in the
case of a nearly defective eigenvalue, eigenvectors becoming nearly linearly
dependent. In the present paper we use orthonormal bases to construct the
appropriate projections associated with the real Schur factorization \cite
{gv} of $f_u(u_0,\lambda )$ and $f_u(u_1,\lambda )$ or their transposes.
This procedure is more accurate and stable.

We first realized the power of the successive continuation approach for
locating connecting orbits when trying to find a homoclinic orbit in a
singular perturbation problem \cite{dfm}. For application to a singularly
perturbed sine-Gordon model of the Josephson junctions and to the
Hodgkin-Huxley equations see \cite{dfg}. The basic idea was also used by
Champneys and Kuznetsov to locate homoclinic orbit in Chua's electronic
circuit and in the FitzHugh-Nagumo equations \cite{ck}. In the context of
optimization, the approach is also used in the tutorial papers \cite{dkk}.

\section{Description of the algorithm}

In addition to the requirement that the fixed points $u_0$ and $u_1$ be
hyperbolic, we assume that the eigenvalues of $f_u(u_0,\lambda )$ and $%
f_u(u_1,\lambda ),$ respectively, satisfy 
\begin{eqnarray}
\func{Re}\mu _{0,n} &\leq &...\leq \func{Re}\mu _{0,n_0+1}<0<\mu _{0,1}<%
\func{Re}\mu _{0,2}\leq ...\leq \func{Re}\mu _{0,n_0},  \tag{2.1a} \\
\func{Re}\mu _{1,1} &\leq &...\leq \func{Re}\mu _{1,n_1}<0<\func{Re}\mu
_{1,n_1+1}\leq ...\leq \func{Re}\mu _{1,n}  \tag{2.1b}
\end{eqnarray}

The method can be extended to the case $\mu _{0,1}=0,$ as in \cite{c, cks}.
It also extends to the cases of complex and multiple $\mu _{0,1}$ by a
modification of Step 0, Eq. (2.13) of the algorithm (see Section 4.3 for an
example). The algorithm requires evaluation of various projections onto
tangent subspaces at $u_0$ and $u_1.$ We construct these projections using
the real Schur factorizations \cite{gv} 
\begin{equation}
f_u(u_0,\lambda )=Q_0T_0Q_0^T,  \tag{2.2}
\end{equation}
and 
\begin{equation}
f_u(u_1,\lambda )=Q_1T_1Q_1^T,  \tag{2.3}
\end{equation}
of $f_u(u_0,\lambda )$ and $f_u(u_1,\lambda ),$ respectively. The
factorization (2.2) has been chosen so that, in the real case,
the eigenvalues $\mu_{0,1},...,\mu _{0,n_0}$ appear in the upper left
corner of $T_0$.
In the complex case the corresponding $2\times 2$ blocks
for each complex congugate pair appear.
A similar choice applies to the eigenvalues $\mu _{1,1},...,\mu _{1,n_1}$ and $%
T_1$ in the factorization (2.3). $Q_0$ $=[q_{0,1}...q_{0,n}]$ and $Q_1$ $%
=[q_{1,1}...q_{1,n}]$ are orthogonal, and $T_0$ and $T_1$ are upper
quasi-triangular (1-by-1 and 2-by-2 blocks on their diagonals). Then the
first $n_0$ columns $q_{0,1},...,q_{0,n}$ of $Q_0$ form an orthonormal basis
of the right invariant subspace $S_0$ of $f_u(u_0,\lambda )$, corresponding
to $\mu _{0,1},...,\mu _{0,n_0}$, and the last $n-n_0$ columns $%
q_{0,n_0+1},...,q_{0,n}$ of $Q_0$ form an orthonormal basis of the
orthogonal complement $S_0^{\bot }$. Similarly, the first $n_1$ columns $%
q_{1,1},...,q_{1,n}$ of $Q_1$ form an orthonormal basis of the right
invariant subspace $S_1$ of $f_u(u_1,\lambda )$, corresponding to $\mu
_{1,1},...,\mu _{1,n_1}$, and the last $n-n_1$ columns $%
q_{1,n_1+1},...,q_{1,n}$ of $Q_1$ form an orthonormal basis of the
orthogonal complement $S_1^{\bot }$.

\newpage
$\mathbf{Approximate}$ $\mathbf{problem}$.

The approximate finite interval problem is now as follows: given $\epsilon
_0=\epsilon _0^{*}\in \Bbb{R}_{+},$ ``small'', $T\in \Bbb{R}_{+},$
``large'', find a solution 
\[
(u^{*},\lambda ^{*},u_0^{*},u_1^{*},d_0^{*},d_1^{*},\epsilon
_1^{*},Q_0^{*},Q_1^{*},T_0^{*},T_1^{*}), 
\]
where $u^{*}\in C^1([0,1]),\Bbb{R}^n)$, $\lambda ^{*}\in \Bbb{R}^{n_\lambda
} $, $u_0^{*},u_1^{*},d_0^{*},d_1^{*},\in \Bbb{R}^n$, $\epsilon _1^{*}$
small, $Q_0^{*},Q_1^{*},T_0^{*},T_1^{*}\in \Bbb{R}^n\times \Bbb{R}^n$, of
the time-scaled differential equation 
\begin{equation}
u^{\prime }(t)-T\text{\/}f(u(t),\lambda )=0,\;0<t<1,  \tag{2.4}
\end{equation}
subject to stationary state conditions 
\begin{eqnarray}
f(u_0,\lambda ) &=&0,  \tag{2.5a} \\
f(u_1,\lambda ) &=&0,  \tag{2.5b}
\end{eqnarray}
left boundary conditions 
\begin{eqnarray}
&&u(0)=u_0+\epsilon _0d_0,  \tag{2.6a} \\
&\mid d_0\mid =1,  \tag{2.6b} \\
&&d_0\cdot q_{0,n_0+j}=0,\;j=1,...,n-n_0,  \tag{2.6c}
\end{eqnarray}
right boundary conditions 
\begin{eqnarray}
&&u(1)=u_1+\epsilon _1d_1,  \tag{2.7a} \\
&\mid d_1\mid =1,  \tag{2.7b} \\
\tau _j &\equiv &d_1\cdot q_{1,n_1+j}=0,\;j=1,...,n-n_1,  \tag{2.8}
\end{eqnarray}
with real Schur factorization of $f_u(u_0,\lambda )$%
\begin{equation}
f_u(u_0,\lambda )q_{0,i}=\sum_{j=1}^nq_{0,j}t_{j,i}^0,\;\;i=1,...,n, 
\tag{2.9a}
\end{equation}
conditions for $Q_0$ to be orthogonal 
\begin{equation}
q_{0,i}^Tq_{0,j}=\delta _{i,j},\;\;\;\;i=1,...,j,\;j=1,...,n,  \tag{2.9b}
\end{equation}
real Schur factorization of $f_u(u_1,\lambda )$%
\begin{eqnarray}
f_u(u_1,\lambda )q_{1,i}=\sum_{j=1}^nq_{1,j}t_{j,i}^1,\;\;i=1,...,n, 
\tag{2.10a}
\end{eqnarray}
and conditions for $Q_1$ to be orthogonal 
\begin{eqnarray}
q_{1,i}^Tq_{1,j}=\delta _{i,j},\;\;\;\;i=1,...,j,\;j=1,...,n.  \tag{2.10b}
\end{eqnarray}
Here the matrix $T_\delta =[t_{j,i}^\delta ],$ $\delta =0,1,$ is upper
quasi-triangular with all elements below the subdiagonal equal to zero.
During each continuation step its $n(n+1)/2$ elements above the subdiagonal
vary, while the remaining $n-1$ elements on the subdiagonal are either fixed
at zero or vary. In the latter case nearby diagonal elements are equated.
Specifically, we have the following: each 1-by-1 block $t_{i,i}^\delta $ on
the diagonal is a real eigenvalue, and each 2-by-2 block on the diagonal $%
\left[ 
\begin{array}{cc}
t_{i,i}^\delta & t_{i,i+1}^\delta \\ 
t_{i+1,i}^\delta & t_{i+1,i+1}^\delta
\end{array}
\right] $ corresponds to a complex conjugate pair, where $t_{i+1,i+1}^\delta
=t_{i,i}^\delta $ is the real part of the eigenvalue and $t_{i,i+1}^\delta
=-t_{i+1,i}^\delta $ is the imaginary part. In the process of continuation
we have the following special cases: (i) if all eigenvalues are real and
distinct then we let all entries on and above the main diagonal vary, while
the entries on the subdiagonal are fixed at value zero; (ii) if we have a
complex conjugate pair then we add the equation $t_{i+1,i+1}^\delta
=t_{i,i}^\delta $ and let $t_{i+1,i}^\delta $ vary, (iii) if we have a
double real eigenvalue then we can choose either of the preceding options.
To summarize, we have 
\begin{eqnarray}
\text{if }t_{i+1,i+1}^0 &\neq &t_{i,i}^0\text{ set }t_{i+1,i}^0=0\text{,
else set }t_{i+1,i+1}^0=t_{i,i}^0\text{ and vary }t_{i+1,i}^0,  \nonumber \\
i &=&1,...,n-1,  \tag{2.9c}
\end{eqnarray}
and similarly 
\begin{eqnarray}
\text{if }t_{i+1,i+1}^1 &\neq &t_{i,i}^1\text{ set }t_{i+1,i}^1=0\text{,
else set }t_{i+1,i+1}^1=t_{i,i}^1\text{ and vary }t_{i+1,i}^1,  \nonumber \\
i &=&1,...,n-1.  \tag{2.10c}
\end{eqnarray}

We remark that a drawback in the earlier version of our algorithm was its
inability to compute through double eigenvalues: we had to stop at double
real eigenvalues and restart with a newly -defined continuation algorithm.
The present method overcomes this problem: one just fixes appropriate
parameters and frees other appropriate parameters (cf. steps 3 and 4 in
Section 4.2).

The above constitutes $n$ differential equations with $n_c=$ $%
3n^2+7n+2-n_0-n_1$ constraints and $n_v=3n^2+5n+n_\lambda +1$ scalar
variables. Generically, it is necessary that $n_c-n_v=n$, in order for the
system (2.4) - (2.10) to have a unique solution. This gives the relation $%
n_\lambda =n-(n_0+n_1)+1.$

By a slight modification of the analysis in \cite{fd1}, using a version of
the implicit function theorem used, e.g., in Beyn \cite{b1}, it is easy to
show, under appropriate transversality conditions, for $T$ large enough and $%
\epsilon _0^{*}$ small enough, that the system (2.4) - (2.10) has a unique
solution which is a good approximation to the solution of (1.1). Note that
since in the present case we look for only one connecting orbit rather than
a branch of connecting orbits, the dimension of the parameter vector $%
\lambda $ is one less than in \cite{fd1}. We also note that the proof in 
\cite{fd1} is for the approximate problem in the same Banach space as the
exact problem and that the result is independent of the particular
implementation, and hence applies to the algorithmic implementation (2.1) -
(2.6b) in \cite{fd1} as well as to (2.4) - (2.10) in the present paper.

\textbf{Algorithm for locating a connecting orbit.}

The solution to the above system is found via a sequence of homotopies that
locate successive zero intercepts of the $\tau _j$, in 
\begin{equation}
\tau _j-d_1\cdot q_{1,n_1+j}(u_1,\lambda )=0,\;j=1,...,n-n_1,  \tag{2.11}
\end{equation}
(cf. equation (2.8)). In each homotopy step we compute a \textit{branch},
i.e., a one-dimensional manifold, of solutions. For this we must have $%
n_c-n_v=n-1, $ and hence $n_\lambda =n-(n_0+n_1)+2$;~ $n_\lambda \geq 0$.

Let $S_{0,k},\;k=1,...,n_0,$ be the right invariant subspace of $%
f(u_0,\lambda _0)$ corresponding to the eigenvalues $\mu _{0,1},...,\mu
_{0,k}.$ Then the first $k$ columns $q_{0,1},...,q_{0,k}$ of $Q_0$ form an
orthonormal basis of $S_{0,k}.$ Initially we replace (2.6) by the equivalent
equations 
\begin{eqnarray}
&&u(0)=u_0+\epsilon _0\sum_{j=1}^{n_0}c_jq_{0j},  \tag{2.12a} \\
&&\sum_{j=1}^{n_0}c_j^2=1.  \tag{2.12b}
\end{eqnarray}
\textbf{Step 0}. Initialize the problem parameter vector $\lambda $, and set
the algorithm parameters $\epsilon _0$ and $T$ to small, positive values, so
that $u(t)$ is approximately constant on $[0,T]$. Set $d_0=q_{01},$%
\begin{equation}
u(t)=u_0+\epsilon _0\ d_0,\;\;\;\;0\leq t\leq 1,  \tag{2.13}
\end{equation}
$\epsilon _1=\left| u(1)-u_1\right| ,$ $\ d_1=(u(1)-u_1)/\epsilon _1,$ $%
c_1=1,$ and $c_2=...=c_{n_0}=0.$ Here the initial direction of the orbit is
typically chosen along the eigenvector of the weakest unstable eigenvalue, $%
d_0=q_{01}$, which implies (2.6), i.e. $u(0)-u_0\in S_0$\textbf{. }\newline
\textbf{Step 1}\textit{. } Compute a solution branch to the system (2.4),
(2.7), (2.11), (2.12a), in the direction of increasing $T$, until $u(1)$
reaches an $\epsilon _1$-neighborhood of $u_1,$ for some $\epsilon _1>0.$
Scalar variables are $T,$ $\epsilon _1\in \Bbb{R},\;d_1\in \Bbb{R}^n,$
$\tau \in \Bbb{R}^{n-n_1}$. There are $n$ differential equations with $n_c=$ 
$3n-n_1+1 $ constraints and $n_v=2n-n_1+2$ scalar variables, and hence $%
n_c-n_v=n-1$. This initial solution is normally a very inaccurate
approximation of the solution of (2.4) - (2.10) since, in general, the $\tau
_j$ in (2.11) will be nonzero and hence $u(1)-u_1\notin S_1.$ In practice
one typically continues until $\epsilon _1$ stops decreasing. Its value is
then not necessarily small, but the successive continuation procedure is
intended to work even then (see also Section~5).
\newline
\textbf{Step }$\mathbf{2}$\textbf{\ (for }$\mathbf{n}_0\mathbf{>1}$\textbf{\
), }$\mathbf{k=2}$\textit{. } Keep $T$ fixed and free $c_1$ and $c_2$, i.e.,
we let $u(0),$ which had been confined to $S_{0,1}$ in Step $1$, now run
through $S_{0,2}$. More precisely, we compute a branch of solutions to the
system (2.4), (2.7), (2.11), (2.12) to locate a zero of, say, $\tau _1$.
Free scalar variables are $\epsilon _1,c_1,c_2\in \Bbb{R},\;d_1\in \Bbb{R}%
^n, $ $\tau \in \Bbb{R}^{n-n_1}$. There are $n$ differential equations with $%
n_c= $ $3n-n_1+2$ constraints and $n_v=2n-n_1+3$ scalar variables, and hence 
$n_c-n_v=n-1$.

\textbf{\ \ \ \ (for }$\mathbf{n}_0\mathbf{>2}$\textbf{\ ), }$\mathbf{%
k=3,...,n}_0$\textit{. } Keep $T$ and $\tau _j=0,\;j=1,...k-2,$ fixed. Also
fix one more component of $\tau $, say $\tau _{k-2}$, which reached value
zero in the previous step. Thus we now let $u(0)-u_0,$ which had been
confined to $S_{0,k-1},$ run through $S_{0,k}.$ More precisely, we compute a
branch of solutions to the system (2.4), (2.7), (2.11), (2.12) to locate
zero of, say, $\tau _{k-1}$. Free scalar variables are $\epsilon
_1,c_1,...,c_k,\tau _{k-1},...,\tau _{n-n_1}\in \Bbb{R},\;d_1\in \Bbb{R}^n$.
There are $n$ differential equations with $n_c=$ $3n-n_1+2$ constraints and $%
n_v=2n-n_1+3$ scalar variables, and hence $n_c-n_v=n-1$.

In the following steps, $\lambda $ also varies. 
For the purpose of implementation, (2.12) is better than (2.6), but the
latter is more stable when eigenvalues coalesce. There are examples where
continuous dependence of $q_{0j}$ on $\lambda $ is lost, and where $c_j$
oscillates. Consequently, starting with Step 3, we use (2.6) instead
of (2.12), since (2.6) is more stable in this case. \newline
\textbf{Step }$\mathbf{3}$\textbf{, }$\mathbf{k=n}_0\mathbf{+1,...,n}_0%
\mathbf{+n}_\lambda \mathbf{\equiv n-n}_1\mathbf{+1}$\textbf{.}\textit{\ }%
The free parameters\textit{\ }$c_1,...,c_k\in \Bbb{R}$ are replaced by $%
d_0\in \Bbb{R}^n.$ Fix another component of $\tau $, say, $\tau _{k-2},$
which reached zero in the previous step, and free one more component of $%
\lambda ,$ say, $\lambda _{k-n_0}$. More precisely, we compute a branch of
solutions to the system (2.4) - (2.7), (2.9) - (2.11). There are $n$
differential equations with $n_c=$ $3n^2+7n+2-n_0-n_1$ constraints and $%
n_v=3n^2+6n+3-n_0-n_1$ scalar variables, and hence $n_c-n_v=n-1.$

In some of the examples below we use slight variations on the basic
algorithm in order to minimize the number of steps, which attests to its
flexibility.

The above algorithm corresponds to Step (II) in the outline of the full
algorithm for locating and continuing connecting orbits in Section~1. Step
(IV) of that outline can now be described as follows.

\textbf{Algorithm for continuing a connecting orbit.}

Compute a branch of solutions to the system (2.4) - (2.10), where $T,$ and
all components of $\lambda $ vary. Here $u(0)\in S_0,$ $u(1)\in S_1,$ $T,$
and all components of $\lambda $ vary. A phase condition 
\begin{equation}
\int_0^1(u^{^{\prime }}(t)-q^{^{\prime }}(t))~\cdot ~u^{^{\prime \prime
}}(t)~dt=0  \tag{2.14}
\end{equation}
may be added if $T$ is kept fixed and $\epsilon _0$ and $\epsilon _1$ are
allowed to vary. Here $q(t)$ is a previously computed orbit on the branch.

\begin{remark}
The factorizations (2.2), (2.3) can be computed at each continuation step
using, for example, LAPACK, or by continuation. Continuation appears to be
more robust. This is to be expected when the eigenvectors, and hence the
Schur matrix, vary rapidly with a slow change of a continuation parameter,
since in this case the pseudo-arclength step will automatically decrease to
capture the transition. We also compared computing a connecting orbit via
continuation and via factorizations in cases where the Schur matrices depend
``smoothly'' on the problem parameters. Computations using continuation were
10-20 times faster (due to larger continuation steps).
\end{remark}

\begin{remark}
For the convenience of the reader, we summarize here some of the issues
related to the algorithms for locating and continuing a connecting orbit, in
this paper and in our earlier work. (1) The \textbf{Approximate problem} in
here is to\textit{\ locate a connecting orbit}, while the approximate
problem in \cite{fd1} is to \textit{continue a branch of connecting orbits}.
Hence the difference between two formulations. (2) Once a connecting orbit
has been located, there are two basic methods to continue a branch of
connecting orbits (see \textbf{Algorithm for continuing a connecting orbit}
on p. 7): (a) $T,$ and $\lambda $ vary, while $\epsilon _0$ and $\epsilon _1$
are kept fixed; (b) $T$ is kept fixed, $\epsilon _0$ and $\epsilon _1$ are
allowed to vary, and a phase condition is added. Both types of continuation
have their advantages and disadvantages. The advantage of (a) is that the
accuracy of the computed orbits on the branch is the same (since $\epsilon
_0 $ and $\epsilon _1$ are fixed), while the advantage of (b) is that it
often works (and works well) for more difficult cases (i.e. when the orbit
changes rapidly during the continuation), when (a) fails or becomes
extremely slow. See also the Remark in \cite{df} which explains (b) in more
detail.
\end{remark}

\begin{remark}
\textbf{An alternate formulation. }An alternate formulation using transposed
matrices, as in \cite{b1, b2, ck, lk}, results in a similar algorithm. The
real Schur factorizations (2.2) and (2.3) are replaced by 
\begin{equation}
f_u^T(u_0,\lambda )=Q_0T_0Q_0^T  \tag{2.15}
\end{equation}
and 
\begin{equation}
f_u^T(u_1,\lambda )=Q_1T_1Q_1^T,  \tag{2.16}
\end{equation}
respectively. Correspondingly, the system (2.4) - (2.10) is replaced by the
system (2.4) - (2.6a,b), (2.7), and 
\begin{equation}
d_0\cdot q_{0,j}=0,\;j=1,...,n-n_0,  \tag{2.17}
\end{equation}

\begin{equation}
\tau _j\equiv d_1\cdot q_{1,j}=0,\;j=1,...,n-n_1,  \tag{2.18}
\end{equation}
\begin{eqnarray}
f_u^T(u_0,\lambda )q_{0,i} &=&\sum_{j=1}^nq_{0,j}t_{j,i}^0,\;\;i=1,...,n, 
\tag{2.19a} \\
\text{if }t_{i+1,i+1}^0 &\neq &t_{i,i}^0\text{\textrm{\ then} set }%
t_{i+1,i}^0=0\text{, else set }t_{i+1,i+1}^0=t_{i,i}^0,\mathrm{and}\text{
vary }t_{i+1,i}^0,  \nonumber \\
i &=&1,...,n-1,  \tag{2.19b} \\
q_{0,i}^Tq_{0,j} &=&\delta _{i,j},\;\;\;\;\;i=1,...,j,\;j=1,...,n, 
\tag{2.19c}
\end{eqnarray}
\begin{eqnarray}
f_u^T(u_1,\lambda )q_{1,i}
&=&\sum_{j=1}^{n-n_1}q_{1,j}t_{j,i}^1,\;\;i=1,...,n,  \tag{2.20a} \\
\text{if }t_{i+1,i+1}^1 &\neq &t_{i,i}^1\text{\textrm{\ then} set }%
t_{i+1,i}^1=0\text{, else set }t_{i+1,i+1}^1=t_{i,i}^1,\mathrm{and}\text{
vary }t_{i+1,i}^1,  \nonumber \\
i &=&1,...,n-1,  \tag{2.20b} \\
q_{1,i}^Tq_{1,j} &=&\delta _{i,j},\;\;\;\;i=1,...,j,\text{ }j=1,...,n. 
\tag{2.20c}
\end{eqnarray}
\end{remark}

\section{Convergence of the algorithm}

The following notation will be used throughout this section: 
\begin{eqnarray}
n_\lambda &=&n+1-n_0-n_1,  \tag{3.1a} \\
n_\tau &=&n-n_1=n_\lambda +n_0-1.  \tag{3.1b}
\end{eqnarray}

In the system (2.4)-(2.10), equations (2.9)-(2.10) can be used to find
numerically $Q_0$ and $Q_1$ as functions of $u_0,\lambda $ and $u_1,\lambda
, $ respectively. This allows to treat (2.4)-(2.8) as a self-contained
problem. Replace the problem (2.4)--(2.8) by an equivalent one 
\begin{equation}
u^{\prime }(t)-T\text{\/}f(u(t),\lambda )=0,\;\;0<t<1,  \tag{3.2}
\end{equation}
\begin{eqnarray}
f(u_0,\lambda ) &=&0,  \tag{3.3a} \\
f(u_1,\lambda ) &=&0,  \tag{3.3b}
\end{eqnarray}
\begin{equation}
\left| u(0)-u_0\right| =\epsilon _0,  \tag{3.4}
\end{equation}
\begin{eqnarray}
\sigma _i &\equiv &(u(0)-u_0)\cdot q_{0,n_0+i}(u_0,\lambda
)=0,\;\;i=1,...,n-n_0,  \tag{3.5a} \\
\sigma _{n-n_0+i} &\equiv &\left( u\left( 0\right) -u_0\right) \cdot
q_{0,i+1}(u_0,\lambda )-\nu _i=0,\;\;i=1,...,n_0-1,  \tag{3.5b}
\end{eqnarray}
\begin{equation}
\tau _i\equiv (u(1)-u_1)\cdot q_{1,n_1+i}(u_1,\lambda )=0,\;\;i=1,...,n_\tau
.  \tag{3.6}
\end{equation}
The equivalence is apparent, since the system (3.2)--(3.4), (3.5a), (3.6)
becomes formally equivalent to the system (2.4)--(2.8) if appended by the
equations 
\begin{eqnarray}
d_0 &=&\left( u\left( 0\right) -u_0\right) /\epsilon _0,  \tag{3.7a} \\
\epsilon _1 &=&\left| u(1)-u_1\right| ,  \tag{3.7b} \\
d_1 &=&(u(1)-u_1)/\epsilon _1,  \tag{3.7c}
\end{eqnarray}
whereas (3.5b) merely defines the parameters $\nu _i.$ Notice that $\nu
_i=c_{i+1},$ $i=1,...,n_0-1,$ where $c_i$'s appear in (2.12).

\textbf{Definition }\textit{A set }$\mathcal{A}$\textit{\ in a Banach space }%
$X$\textit{\ will be called an arc if there exists a Banach space }$Y$%
\textit{\ and an open sets }$U\subset X,$\textit{\ }$V\subset Y$\textit{\
and a }$C^1$\textit{-diffeomorphism }$F:U\rightarrow V$\textit{\ such that }$%
\mathcal{A}\subset U$\textit{\ and }$F(\mathcal{A})$\textit{\ is a closed
bounded rectilinear segment (a degenerate segment consisting of a single
point is allowed).}

\begin{remark}
Any arc $\mathcal{A}$ has finite length. Indeed, if $\gamma
:[0,1]\rightarrow Y$ is a linear parametrization of the rectilinear segment $%
F(\mathcal{A}),$ then $F^{-1}\circ \gamma $ is a smooth parametrization of $%
\mathcal{A}$ with continuous, hence bounded, velocity $v(s)=\left[
F^{-1}\right] ^{\prime }(\gamma (s))\circ \gamma ^{\prime }(s).$ Thus $%
(length$ $of$ $A)\leq $ $\sup \left| v(s)\right| .$
\end{remark}

Introduce two Banach spaces, $X=C^1[0,1]\times \Bbb{R}^n\times \Bbb{R}%
^n\times \Bbb{R}^{n_\tau }=\{x=(U,\Lambda ):U=(u,u_0,u_1)\in C^1[0,1]\times 
\Bbb{R}^n\times \Bbb{R}^n,\;u\in C^1[0,1],u_0,u_1\in \Bbb{R}^n,\;\Lambda
=(\nu ,\lambda )\in \Bbb{R}^{n_\tau },\;\nu \in \Bbb{R}^{n_0-1},\;\lambda
\in \Bbb{R}^{n_\lambda }\}$ and $Y=C[0,1]\times \Bbb{R}^n\times \Bbb{R}%
^n\times \Bbb{R}\times \Bbb{R}^{n-1}.$ Define $F:X\rightarrow Y$ by 
\begin{equation}
F(U,\Lambda )=\left( u^{\prime }-T\,f(u,\lambda ),\;f(u_0,\lambda
),\;f(u_1,\lambda ),\;\left| u(0)-u_0\right| ^2-\epsilon _0^2,\;\sigma
(U,\Lambda )\right) ,  \tag{3.8}
\end{equation}
where $\sigma :X\rightarrow \Bbb{R}^{n-1},\;\sigma =(\sigma _1,...,\sigma
_{n-1})$ is defined by 
\begin{eqnarray}
\sigma _i(U,\Lambda ) &=&(u(0)-u_0)\cdot q_{0,n_0+i}(u_0,\lambda
),\;\;i=1,...,n-n_0  \nonumber \\
&&  \tag{3.9a} \\
\sigma _i(U,\Lambda ) &=&\left( u\left( 0\right) -u_0\right) \cdot
q_{0,i+1-(n-n_0)}(u_0,\lambda )-\nu _{i-(n-n_0)},\;i=n-n_0+1,...,n-1. 
\nonumber \\
&&  \tag{3.9b}
\end{eqnarray}
Also define $\tau :X\rightarrow \Bbb{R}^{n_\tau },\;\tau =(\tau _1,...,\tau
_{n_\tau })$ by 
\begin{equation}
\tau _j(U,\Lambda )=(u(1)-u_1)\cdot q_{1,n_1+j}(u_1,\lambda
),\;\;j=1,...,n_\tau .  \tag{3.10}
\end{equation}

\textbf{Theorem }\textit{Define }$F^{(j)}:X\rightarrow Y\times \Bbb{R}%
^{n_\tau },\;F^{(j)}=\left( F_Y^{(j)},F_1^{(j)},...,F_{n_\tau }^{(j)}\right)
,\;j=1,...,n_\tau ,$\textit{\ by } 
\begin{eqnarray}
F^{(j)}(U,\Lambda ) &=&(F(U,\Lambda ),\tau _1(U,\Lambda ),...,\tau
_j(U,\Lambda ),\Lambda _{j+1},...,\Lambda _{n_\tau }),\;j=1,...,n_\tau -1, 
\nonumber \\
&&  \tag{3.11a} \\
F^{(n_\tau )}(U,\Lambda ) &=&(F(U,\Lambda ),\tau (U,\Lambda )).  \nonumber \\
&&  \tag{3.11b}
\end{eqnarray}
\textit{Assume that there exist }$(U^{*},\Lambda ^{*})\in X$\textit{\ such
that } 
\begin{equation}
F^{(n_\tau )}(U^{*},\Lambda ^{*})=0.  \tag{3.12}
\end{equation}
\textit{Assume also that for }$j=1,...,n_\tau ,\;F^{(j)}$\textit{\ is a }$%
C^1 $\textit{-mapping\ in an open neighborhood of }$(U^{*},\Lambda ^{*})$%
\textit{, and its Frech\'{e}t derivative at }$(U^{*},\Lambda ^{*})$\textit{\
is an invertible bounded linear operator.}

\textit{Then there exists an open neighborhood }$V$\textit{\ of }$%
(U^{*},\Lambda ^{*})$\textit{\ such that for any }$(U^0,\Lambda ^0)$\textit{%
\ }$\in V\cap \{(U,\Lambda ):F(U,\Lambda )=0\}$\textit{\ there exists a
continuous piecewise smooth curve }$x(s)=(U(s),\Lambda (s)),\;s\in
[0,1],\;x(0)=(U^0,\Lambda ^0),$\textit{\ }$x(1)=(U^{*},\Lambda ^{*})$\textit{%
\ with the following properties:}

\textit{(i) } $x([0,1])\subset V$\textit{\ and consists of } $n_\tau $%
\textit{\ arcs, specifically, there exist numbers } $0\leq s_0\leq s_1\leq
..\leq s_{n_\tau }=1$ such that each $\mathcal{A}%
_k=x([s_{k-1},s_k]),k=1,...,n_\tau $ is an arc;

\textit{(ii) for any } $s\in [0,1],$ \textit{\ }$F(x(s))=0$ \textit{;}

\textit{(iii) for any }$k\in \{2,...,n_\tau \},$\textit{\ if }$x\in \mathcal{%
A}_k,$ \textit{then } $\tau _1(x)=...=\tau _{k-1}(x)=0;$

\textit{(iv) for any } $s\in [0,1],$ \textit{if }$\tau _k(x(s))\neq 0$ 
\textit{\ for some }$k\in \{1,...,n_\tau -1\},$\textit{\ then }$\Lambda
_{k+1}(s)=\Lambda _{k+1}^0,...,\Lambda _{n_\tau }(s)=\Lambda _{n_\tau }^0.$
\newline
The hypotheses of the Theorem which require that \textit{Frech\'{e}t
derivatives of} $F^{(j)}$\textit{\ at the solution }$(U^{*},\Lambda ^{*})$%
\textit{\ are invertible bounded linear operators,} express the following
transversality requirements. The surface $\{F=0\}$ and the hypersurfaces $%
\{\tau _1=0\},...,\{\tau _{n_\tau }=0\}$ intersect transversely at the
solution $(U^{*},\Lambda ^{*})$; the same holds for the surface $\{F=0\}$
and the hypersurfaces $\{\tau _1=0\}$,...,$\{\tau _j=0\}$,$\{\Lambda
_{j+1}=\Lambda _{j+1}^{*}\}$,...,$\{\Lambda _{n_\tau }=\Lambda _{n_\tau
}^{*}\},$ for each $j=1,...,n_\tau -1$. In particular, these requirements
imply that, at the end of each step $j$ $(j=1,...,n_\tau )$, $\tau _j$
crosses zero transversely.

Define 
\begin{equation}
\Sigma _0=\{x\in X:F(x)=0\},\;\;\Sigma _i=\Sigma _0\cap
\bigcap_{k=1}^i\{x\in X:\tau _k(x)=0\},\;i=1,...,n_\tau .  \tag{3.13}
\end{equation}

The following Lemma 1 may be interpreted as Lemma 2$_j$ below for $j=n_\tau
. $ To emphasize this, the condition (iv) is included in Lemma 1 even though
it is a trivial consequence of (iii)$.$ Incorporation of the case $j=n_\tau $
into the statement of Lemma 2$_j$ would lead to a cumbersome formulation.

\textbf{Lemma 1 } \textit{Let }$F,$\textit{\ }$\tau ,$\textit{\ }$U^{*},$%
\textit{\ }$\Lambda ^{*},$\textit{\ }$F^{(n_\tau )}$\textit{\ satisfy the
hypothesis of the Theorem. Then there exists an open neighborhood }$W$%
\textit{\ of }$(U^{*},\Lambda ^{*})$\textit{\ such that for any }$(U^{n_\tau
-1},\Lambda ^{n_\tau -1})\in W\cap \Sigma _{n_\tau -1}$\textit{\ there
exists a smooth curve }$x(s)=(U(s),\Lambda (s))$\textit{, }$s\in [0,1],$%
\textit{\ }$x(0)=(U^{n_\tau -1},\Lambda ^{n_\tau -1}),\;x(1)=(U^{*},\Lambda
^{*})$\textit{\ with the following properties:}

\textit{(i) }$x([0,1])\subset W$\textit{\ and is an arc;}

\textit{(ii) for any }$s\in [0,1],$\textit{\ }$F(x(s))=0$\textit{;}

\textit{(iii) for any }$k\in \{1,...,n_\tau -1\}$\textit{, }$\tau _k(x(s))=0$%
\textit{\ for all }$s\in [0,1].$

\textit{(iv) for any }$s\in [0,1],$\textit{\ if }$\tau _k(x(s))\neq 0$%
\textit{\ for some }$k\in \{1,...,n_\tau -1\},$\textit{\ then }$\Lambda
_{k+1}(s)=\Lambda _{k+1}^{n_\tau -1},...,\Lambda _{n_\tau }(s)=\Lambda
_{n_\tau }^{n_\tau -1};$

\textbf{Lemma 2}$_j$ \textit{Let\ }$j\in \{1,...,n_\tau -1\}$\textit{\ and
let }$F,$\textit{\ } $\tau ,$ \textit{\ }$U^{*},$\textit{\ }$\Lambda ^{*},$%
\textit{\ }$F^{(j)}$ \textit{\ satisfy the hypothesis of the Theorem. Assume
that there exists an open neighborhood } $W^{(j)}$ \textit{\ of } $%
(U^{*},\Lambda ^{*})$ \textit{\ such that for any } $(U^j,\Lambda ^j)\in
W^{(j)}\cap \Sigma _j$ \textit{\ there exists a continuous piecewise smooth
curve } $x^{(j)}(s)=(U^{(j)}(s),\Lambda ^{(j)}(s)),\;s\in
[0,1],\;x^{(j)}(0)=(U^j,\Lambda ^j),\;x^{(j)}(1)=(U^{*},\Lambda ^{*})$%
\textit{\ with the following properties:}

\textit{(i}$_j$\textit{) } $x^{(j)}([0,1])\subset W^{(j)}$ \textit{\ and
consists of } $n_\tau -j$ \textit{\ arcs, specifically, there exist numbers }
$0\leq s_j^{(j)}\leq s_{j+1}^{(j)}\leq ...\leq s_{n_\tau }^{(j)}=1$ such
that each $\mathcal{A}_k^{(j)}=x^{(j)}([s_{k-1}^{(j)},s_k^{(j)}]),$ $%
k=j+1,...,n_\tau $ is an arc;

\textit{(ii}$_j$\textit{)\ for any } $s\in [0,1],\;F\left( x^{(j)}(s)\right)
=0;$

\textit{(iii}$_j$\textit{) for any }$k\in \{j+1,...,n_\tau \},$\textit{\ if }%
$x\in \mathcal{A}_k^{(j)},$ \textit{then }$\tau _1(x)=...=\tau _{k-1}(x)=0.$

\textit{(iv}$_j$\textit{) for any }$s\in [0,1],$\textit{\ if }$\tau _k\left(
x^{(j)}(s)\right) \neq 0$\textit{\ for some }$k\in \{1,...,n_\tau -1\},$%
\textit{\ then }$\Lambda _{k+1}(s)=\Lambda _{k+1}^j,...,\Lambda _{n_\tau
}(s)=\Lambda _{n_\tau }^j.$

\textit{Then there exists an open neighborhood }$W^{(j-1)}$\textit{\ of }$%
(U^{*},\Lambda ^{*})$\textit{\ such that for any }$(U^{j-1},\Lambda
^{j-1})\in W^{(j-1)}\cap \Sigma _{j-1}$\textit{\ there exists a continuous
piecewise smooth curve }$x^{(j-1)}(s),$\textit{\ }$s\in
[0,1],\;x^{(j-1)}(0)=(U^{j-1},\Lambda ^{j-1}),\;x^{(j-1)}(1)=(U^{*},\Lambda
^{*})$\textit{\ which satisfies the conditions }$(i)-(iv)$\textit{\ above
with }$j$\textit{\ substituted by }$j-1.$

\textbf{Proof of Theorem.} Lemma 1 guarantees the existence of a
neighborhood described in the hypothesis of Lemma 2$_{n_\tau -1}$ (take $%
W^{(n_\tau -1)}=W).$ The conclusion of the latter assures the existence of a
neighborhood described in the assumption of Lemma $2_{n_\tau -2},$ and so
on. At the end one finds that there exists a neighborhood $W^{(0)}$
described in the conclusion of Lemma $2_1.$ The properties of $W^{(0)}$
coincide with the requirements for $V$. Thus take $V=W^{(0)}.$

\textbf{Proof of Lemma 1}\QTR{textbf}{.} By the Inverse Mapping Theorem \cite
{d, l}, applied to $F^{(n_\tau )}$ there exist open neighborhoods $A,$ $%
A^{\prime }$of $(U^{*},\Lambda ^{*})$ and $F^{(n_\tau )}(U^{*},\Lambda
^{*})=0,$ respectively, such that the restriction $F^{(n_\tau
)}:A\rightarrow A^{\prime }$ has a continuously differentiable inverse $%
H:A^{\prime }\rightarrow A.$ Let $B$ be an open ball in $Y\times \Bbb{R}%
^{n_\tau }$ centered at $0$ and contained in $A^{\prime }.$ Define $W=H(B).$
Let $(U^{n_\tau -1},\Lambda ^{n_\tau -1})\in W\cap \Sigma _{n_\tau -1}.$
Define 
\begin{equation}
x(s)=H(y(s)),\;s\in [0,1],  \tag{3.14}
\end{equation}
where 
\begin{equation}
y(s)=(y_Y(s),y_1(s),...,y_{n_\tau }(s))=(0_{Y,}0,...,0,(1-s)\tau _{n_\tau
}(U^{n_\tau -1},\Lambda ^{n_\tau -1})),  \tag{3.15}
\end{equation}
and $0_Y$ is the zero element in $Y.$

To see that $x$ satisfies condition \QTR{textit}{(i)}, notice that $%
y(s),\;s\in [0,1],$ is a rectilinear segment connecting $y(0)=F^{(n_\tau
)}(U^{n_\tau -1},\Lambda ^{n_\tau -1})$ to $y(1)=0.$ The latter is the
center of the ball $B$ and $F^{(n_\tau )}(U^{n_\tau -1},\Lambda ^{n_\tau
-1})\in B.$ Therefore $y[0,1]\subset B$ (so, in particular, $H(y(s)),\;s\in
[0,1],$ is well defined). We conclude that $x([0,1])=H(y([0,1]))$is an arc.

That $x(s)$ satisfies condition \QTR{textit}{(ii)} is clear from $%
F(x(s))=F_Y^{(n_\tau )}(x(s))=y_Y(s)=0,\;s\in [0,1].$

Notice that, for $k=1,...,n_\tau -1,\;\tau _k(x(s))=F_k^{(n_\tau
)}(x(s))=y_k(s)=0$ for all $s\in [0,1],$ i.e. condition (iii) holds.

Condition \QTR{textit}{(iv)} is a trivial consequence of (iii)$.$

\textbf{Proof of Lemma 2}$_j$\textbf{.} By the Inverse Mapping Theorem
applied to $F^{(j)}$ there exist open neighborhoods $A,$ $A^{\prime },$ of $%
(U^{*},\Lambda ^{*})$ and of $y^{*}=F^{(j)}(U^{*},\Lambda ^{*})=\left(
0_Y,0,...,0,\Lambda _{j+1}^{*},...,\Lambda _{n_\tau }^{*}\right) ,$
respectively, such that $F^{(j)}:A\rightarrow A^{\prime }$ has a
continuously differentiable inverse $H:A^{\prime }\rightarrow A.$ Then $%
F^{(j)}\left( W^{(j)}\cap A\right) =H^{-1}\left( W^{(j)}\cap A\right) $ is
open and contains $y^{*}$ (since both $W^{(j)}$ and $A$ are open, both
contain $(U^{*},\Lambda ^{*}),$ and $H$ is continuous). Therefore there
exists an open ball $B$ in $Y\times \Bbb{R}^{n_\tau }$ centered at $y^{*}$
and contained in $F^{(j)}\left( W^{(j)}\cap A\right) .$ Set $W^{(j-1)}=H(B).$

Let us verify that $W^{(j-1)}$ possesses the desired properties. Let $%
(U^{j-1},\Lambda ^{j-1})\in W^{(j-1)}\cap \Sigma _{j-1}.$ Define a
parametrized curve $x(s)$ by 
\begin{equation}
x(s)=H(y(s)),\;s\in [0,1],  \tag{3.16}
\end{equation}
where 
\begin{equation}
y(s)=\left( y_Y(s),y_1(s),...,y_{n_\tau }(s)\right) =\left(
0_Y,0,...,0,(1-s)\tau _j(U^{j-1},\Lambda ^{j-1}),\Lambda
_{j+1}^{j-1},...,\Lambda _{n_\tau }^{j-1}\right) .  \tag{3.17}
\end{equation}
Notice that $y(0)=F^{(j)}(U^{j-1},\Lambda ^{j-1})\in B$ and 
\begin{eqnarray}
\ &\parallel &y(0)-y^{*}\parallel \,=\max \{\parallel y_Y\parallel
_Y,\parallel (y_1,...,y_{n_\tau })\parallel _{\Bbb{R}^{n_\tau }}\}=\parallel
(y_1,...,y_{n_\tau })\parallel _{\Bbb{R}^{n_\tau }}  \nonumber \\
\ &=&\left( \left[ \tau _j(U^{j-1},\Lambda ^{j-1})\right] ^2+\left( \Lambda
_{j+1}^{j-1}-\Lambda _{j+1}^{*}\right) ^2+...+\left( \Lambda _{n_\tau
}^{j-1}-\Lambda _{n_\tau }^{*}\right) ^2\right) ^{1/2}  \nonumber \\
\ &\geq &\left( \left( \Lambda _{j+1}^{j-1}-\Lambda _{j+1}^{*}\right)
^2+...+\left( \Lambda _{n_\tau }^{j-1}-\Lambda _{n_\tau }^{*}\right)
^2\right) ^{1/2}=\parallel y(1)-y^{*}\parallel ,  \tag{3.18}
\end{eqnarray}
i.e., $y(1)$ is not farther from the center $y^{*}$ of the ball $B$ than $%
y(0).$ Hence $y(1)\in B,$ and so $y([0,1]),$ which is a rectilinear segment,
is contained in $B.$ This implies, first, that $y([0,1])$ is in the domain
of $H,$ hence $x(s)$ is well defined, and, second, that $x([0,1])=H(y[0,1]))%
\subset W^{(j-1)}$ is an arc, since $H$ is a diffeomorphism and $y([0,1])$
is a rectilinear segment.

Notice that $x(1)\in W^{(j)}\cap S_{j.}$ Therefore, by assumption, there
exists a continuous piecewise smooth curve $x^{(j)}(s),$ $s\in [0,1],$ such
that $x^{(j)}(0)=x(1),$ $x^{(j)}(1)=(U^{*},\Lambda ^{*}),$ and the
conditions (i$_{\text{j}}$) - (iv$_{\text{j}}$) are satisfied. Define

\begin{equation}
x^{(j-1)}(s)=\left\{ 
\begin{tabular}{ll}
$x(2s),$ & $0\leq s<1/2,$ \\ 
$x^{(j)}(2s-1),$ & $1/2\leq s\leq 1.$%
\end{tabular}
\right.  \tag{3.19}
\end{equation}

By construction, $x^{(j-1)}(s)$ is continuous, piecewise smooth, and $%
x^{(j-1)}([0,1])\subset W^{(j-1)}.$ Also $x^{(j-1)}([0,1])$ consists of $%
n_\tau -j+1$ arcs, since $x^{(j-1)}([0,\frac 12])=x([0,1])$ is an arc,
whereas $x^{(j-1)}([\frac 12,1])=x^{(j)}([0,1])$ was assumed to consist of $%
n_\tau -j$ arcs. More precisely, if $s_j^{(j)},...,s_{n_\tau }^{(j)}$ are
such as assumed in (iii$_{\text{j}}$)$,$ then we define $s_{j-1}^{(j-1)}=0,$ 
$s_k^{(j-1)}=(s_k^{(j)}+1)/2,$ $k=j,...,n_\tau $ (clearly, $0\leq
s_{j-1}^{(j-1)}\leq s_j^{(j-1)}\leq ...\leq s_{n_\tau }^{(j-1)}=1).$ Then $%
\mathcal{A}_j^{(j-1)}\equiv
x^{(j-1)}([s_{j-1}^{(j-1)},s_j^{(j-1)}])=x^{(j-1)}([0,\frac 12])=x([0,1])$
and, for $k=j+1,...,n_\tau ,$ $\mathcal{A}_k^{(j-1)}\equiv
x^{(j-1)}([s_{k-1}^{(j-1)},s_k^{(j-1)}])=x^{(j-1)}([\frac{s_{k-1}^{(j)}+1}2,%
\frac{s_k^{(j)}+1}2])=x^{(j)}([s_{k-1}^{(j)},s_k^{(j)}])=\mathcal{A}_k^{(j)}$
are arcs. Thus condition (i$_{\text{j-1}}$) is satisfied.

Condition (ii$_{\text{j-1}}$) is satisfied for $x^{(j-1)}(s),$ since it is
satisfied for both $x(s)$ and $x^{(j)}(s).$

If $x\in \mathcal{A}_j^{(j-1)},$ i.e. $x=x(s)$ for some $s\in [0,1],$ then,
for $m=1,...,j-1,$ $\tau
_m(x)=F_m^{(j)}(x)=F_m^{(j)}(x(s))=F_m^{(j)}(H(y(s)))=y_m(s)=0.$ If $x\in 
\mathcal{A}_k^{(j-1)}$ for some\textit{\ }$k\in \{j+1,...,n_\tau \},$ then,
since $\mathcal{A}_k^{(j-1)}=\mathcal{A}_k^{(j)},$ we have\textit{\ }$\tau
_1(x)=...=\tau _{k-1}(x)=0$ by (iii$_{\text{j}}$). Thus (iii$_{\text{j}}$)
holds.

To verify (iv$_{\text{j-1}}$) for $x^{(j-1)}(s),$ one has to consider $s\in
[0,\frac 12)$ only, since, for $s\in [\frac 12,1],$ $%
x^{(j-1)}(s)=x^{(j)}(2s-1),$ and (iv$_{\text{j}}$) applies. If $s\in
[0,\frac 12),$ then $x^{(j-1)}(s)=x(2s)$, so, from the definition of $x(s)$, 
$\tau _1\left( x^{(j-1)}(s)\right) =...=\tau _{j-1}\left(
x^{(j-1)}(s)\right) =0.$ Thus the implication in (iv$_{\text{j-1}}$) has to
be verified for $k\geq j$ only. Suppose that $\tau _j\left(
x^{(j-1)}(0)\right) \equiv \tau _j\left( U^{j-1},\Lambda ^{j-1}\right) \neq
0 $ (otherwise the arc $x([0,1])$ degenerates into a single point $%
x^{(j-1)}(\frac 12),$ which has already been considered. Then, for any $s\in
[0,\frac 12),$%
\begin{equation}
\tau _j(x(2s))=\tau _j(H(y(2s)))=F_j^{(j)}(H(y(2s)))=y_j(2s)=(1-2s)\tau
_j(U^{j-1},\Lambda ^{j-1})\neq 0.  \tag{3.20}
\end{equation}
Let us first verify (iv$_{\text{j-1}}$) for $k=j.$ For any $x\in X,$ its
last $n_\tau -j$ coordinates coincide with those of $F^{(j)}(x)$ by the
definition of $F^{(j)},$ c.f. (3.11a). But $F^{(j)}(x(s))=y(s)$ whose last $%
n_\tau -j$ coordinates are $\Lambda _{j+1}^{j-1},...,\Lambda _{n_\tau
}^{j-1},$ respectively. This proves the implication contained in (iv$_{\text{%
j-1}}$) for $k=j.$

Validity of (iv$_{\text{j-1}}$) for $k>j$ follows immediately from the above
observation that the last $n_\tau -j$ coordinates of $x(s),$ $0\leq s\leq 1,$
are $\Lambda _{j+1}^{j-1},...,\Lambda _{n_\tau }^{j-1},$ respectively.

\begin{remark}
It follows from the construction of the curves $x(s)$ in the proofs of
Lemmas 1 and 2$_j$ that, when $\Lambda _j$ is ``freed,'' the value of $\mid
\tau _j(x(s))\mid $ strictly decreases, as $s$ goes from $0$ to $1,$
provided $\tau _j(x(0))\neq 0.$ This is immediate from $\tau
_j(x(s))=F_j^{(j)}(H(y(s)))=y_j(s)=(1-s)\tau _j(x(0)).$
\end{remark}

\begin{remark}
Curves constructed in the proofs of Lemmas 1 and 2$_j$ have finite length.
Indeed, they consist of a finite number of arcs of finite length, see Remark
4.
\end{remark}

\section{Examples}

\subsection{Example 1: Homoclinic orbits in a 3-D singular perturbation
problem.}

We compute a homoclinic orbit for the 3-D system \cite{d1} 
\begin{eqnarray}
x^{\prime } &=&(2-z)a(x-2)+(z+2)[\alpha (x-x_0)+\beta (y-y_0)],  \tag{4.1a}
\\
y^{\prime } &=&(2-z)[d(b-a)(x-2)/4+by]+(z+2)[-\beta (x-x_0)+\alpha (y-y_0)],
\tag{4.1b} \\
z^{\prime } &=&(4-z^2)[z+2-m(x+2)]-\epsilon cz,  \tag{4.1c}
\end{eqnarray}
where $a=1,$ $b=1.5,$ $c=2,$ $d=-.2,$ $m=1.1845,$ $\alpha =.01,$ $\beta =5,$ 
$x_0=-.1,$ $y_0=-2$. The parameter $\epsilon $ is taken as variable. In this
case $n_0=2$ and $n_1=1$ in Equation (2.1). The discretization is orthogonal
collocation with piecewise polynomials, using 25 subintervals and 4
collocation points per interval. A relative Newton tolerance of $10^{-8}$ is
used for $u$ and $\lambda .$

\textbf{Step 0}. Initialize the problem parameter: $\epsilon =.01$, and the
algorithm parameters: $\epsilon_0=10^{-4}$, $T=10^{-2},$ $c_1=1, $ $c_2=0.$

\textbf{Step 1}\textit{. } Compute a branch of solutions to the equations
(2.4), (2.7), (2.11), (2.12a), and 
\begin{equation}
d_0=\sum_{j=1}^{n_0}c_jq_{0j},  \tag{4.2}
\end{equation}
with scalar variables $T,\epsilon _1\in \Bbb{R},\;d_0,d_1\in \Bbb{R}^3,$ $%
\tau _1,\tau _2\in \Bbb{R}$, in the direction of increasing $T$ until $\tau
_1$ crosses zero. There are ten scalar variables in this continuation. Final
values are $T=2.7568,$ $\epsilon _1=3.509$, and $\tau _2=.0223$.

\textbf{Step 2}$.$ Fix $\tau _1=0$ and free $c_1$ and $c_2$ in (4.3).
Compute a solution branch to the system (2.4), (2.7), (2.11), (2.12), and
(4.3), in the direction of decreasing $c_2$ until $\tau _2$ crosses zero.
There are now eleven free scalar variables. Final values are $c_1=1.0$, $%
c_2=-1.51\times 10^{-4},$ $T=2.7580,$ and $\epsilon _1=2.548.$

\textbf{Step 3}$.$ The free parameters\textit{\ }$c_1,c_2\in \Bbb{R}$ are
replaced by $d_0\in \Bbb{R}^3$; see Equation (2.6). Fix $\tau _1=0$ and free 
$\epsilon \in \Bbb{R}$ and hence the matrices $Q_0,Q_1,T_0,T_1\in \Bbb{R}%
^3\times \Bbb{R}^3.$ Compute a solution branch to (2.4), (2.5a), (2.6),
(2.7), (2.9) - (2.11) in the direction of decreasing $\epsilon _1$ (to
increase the accuracy of the orbit) until $\epsilon _1=\epsilon _0$. There
are 44 free scalar variables. Terminal values are $\epsilon =.009333142,$ $%
T=2.7679,$ and $\epsilon _1=10^{-4}.$

Note, that it took only 2 steps to get $\tau _1=\tau _2=0,$ while the
general algorithm requires 3 steps to accomplish this. In particular, since $%
\tau _1$ crossed zero already in the first step, we were able (without
changing the total number of free scalar variables) to keep $T$ free in the
second step, while $\tau _1=0$ was fixed.

\subsection{Example 2: Heteroclinic orbits in a 3-D Josephson Junction
problem.}

A singularly perturbed sine-Gordon equation, modeling magnetic flux quanta
(``fluxons'') in long Josephson tunnel junctions with nonzero surface
impedance, 
\begin{equation}
\beta c\phi ^{\prime \prime \prime }(\xi )-(1-c^2)\phi ^{\prime \prime }(\xi
)-\alpha c\phi ^{\prime }(\xi )+\sin \phi (\xi )-\gamma =0,  \tag{4.3}
\end{equation}
was studied by several authors, see e.g. \cite{bfmp} and references therein.
In \cite{dfg} we computed single and multiple fluxon solutions, which are
heteroclinic orbits (or homoclinic orbits on a cylinder). The algorithm in 
\cite{dfg} does not allow computation of the transition from two real
eigenvalues to a complex conjugate pair. Here we accurately compute this
transition point. The three-dimensional first order system is 
\begin{eqnarray}
\phi _1^{\prime } &=&\phi _2,  \tag{4.4a} \\
\phi _2^{\prime } &=&\phi _3,  \tag{4.4b} \\
\phi _3^{\prime } &=&[(1-c^2)\phi _3+\alpha c\phi _2-\sin \phi _1+\gamma
]/\beta c.  \tag{4.4c}
\end{eqnarray}
We compute a heteroclinic orbit with $u_0=(\arcsin \gamma ,0,0)$and $%
u_1=(\arcsin \gamma +\pi ,0,0).$ Throughout, $\alpha =.18$ and $\beta =.1$
are kept fixed, and $\gamma $ and $c$ vary. In this case $n_0=1$ and $n_1=2$
in Equation (2.1). Discretization is as in the preceding example.

\textbf{Step 0.} Initialize the problem parameters, $\gamma =.1$, $c=.6$,
and the algorithm parameters, $\epsilon _0=10^{-4}$, $\epsilon _1=.6283$, $%
T=10^{-2},$ and $c_1=1.$

\textbf{Step 1.}\QTR{textit}{\ } Compute a solution branch to (2.4), (2.7),
(2.11), (2.12a), (4.2), with scalar variables $T,\epsilon _1\in \Bbb{R}%
,\;d_0,d_1\in \Bbb{R}^3,$$\tau \in \Bbb{R}$, in the direction of increasing $%
T$ until $\tau $ crosses zero. There are nine free scalar variables.
Terminal values are $T=9.336,$ and $\epsilon _1=3.642.$

\textbf{Step 2}$.$ Fix $\tau _1=0$. Free $c\in \Bbb{R}$, and hence the
matrices $Q_0,Q_1\in \Bbb{R}^3\times \Bbb{R}^3$, and the entries $%
t_{11}^0,t_{12}^0,t_{13}^0,t_{22}^0,t_{23}^0,t_{33}^0$ and $%
t_{11}^1,t_{12}^1,t_{13}^1,t_{22}^1,t_{23}^1,t_{33}^1$ of the matrices $%
T_0,T_1\in \Bbb{R}^3\times \Bbb{R}^3,$ respectively, as well as $t_1,$$t_2, $
in Equation (4.5) below. Compute a branch of solutions to the system (2.4),
(2.5a), (2.6) - (2.10), and 
\begin{eqnarray}
t_1 &=&t_{22}^0-t_{33}^0,  \tag{4.5a} \\
t_2 &=&t_{11}^1-t_{22}^1,  \tag{4.5b}
\end{eqnarray}
in the direction of decreasing $\epsilon _1,$ until $\epsilon _1=\epsilon _0$%
. This step involves 44 free scalar variables. Terminal values are $c=.3404,$
$T=21.13,$ $\epsilon _1=10^{-4},\;\mu _{01}=1.073,$ $\mu _{11}=-2.600$, and $%
\mu _{12}=-1.047.$ Above, $t_{22}^0=\func{Re}\mu _{02},$ $t_{33}^0=\func{Re}%
\mu _{03},$ $t_{11}^1=\func{Re}\mu _{11},$ $t_{22}^1=\func{Re}\mu _{12}.$
The parameters $t_1,$$t_2$ in (4.5) serve as test functions that cross zero
when $\mu _{02}$ and $\mu _{03}$ (and $\mu _{11}$ and $\mu _{12})$ are
multiple.

\textbf{Step 3}$.$ Fix $\epsilon _1$ and free $\gamma \in \Bbb{R}$. Compute
a solution branch with parameters $\gamma $ and $c$ in the direction of
decreasing $\gamma $ until $t_1$ and $t_2$ cross zero. The equations are as
in Step~2. Terminal values are $\gamma =4790,$ $c=.3404,$ $T=12.92,$ $\mu
_{01}=1.622,$ and $\mu _{11}=\mu _{12}=-2.534.$ The double eigenvalue was
located with accuracy $10^{-8}$.

\textbf{Step 4}$.$ To continue the complex conjugate pair of the
eigenvalues, fix $t_1=t_2=0$ $(\func{Re}\mu _{11}=\func{Re}\mu _{12})$ and
free $t_{32}^0,t_{21}^1$ ($\func{Im}\mu _{02},$$\func{Im}\mu _{12}$).
Compute a solution branch with parameters $\gamma $ and $c$ to the same
system as in Step 3 in the direction of decreasing $\gamma .$ Final values
are $\gamma =0.8830$, $c=1.000$, and $T=30.67$.

Note, that it took only one step to get $\tau =0,$ while the general
algorithm requires two steps to accomplish this.

\subsection{Example 3: Heteroclinic orbits in a modified 3-D Josephson
Junction problem.}

Consider system (4.4) with reversed time, 
\begin{eqnarray}
\phi _1^{\prime } &=&-\phi _2,  \tag{4.6a} \\
\phi _2^{\prime } &=&-\phi _3,  \tag{4.6b} \\
\phi _3^{\prime } &=&-[(1-c^2)\phi _3+\alpha c\phi _2-\sin \phi _1+\gamma
]/\beta c.  \tag{4.6c}
\end{eqnarray}
We compute a heteroclinic orbit with $u_0=(\arcsin \gamma +\pi ,0,0)$ and $%
u_1=(\arcsin \gamma ,0,0).$ As before, $\alpha =.18$ and $\beta =.1$ are
kept fixed, and $\gamma $ and $c$ vary. The problem parameters are chosen so
that $n_0=2$ and $n_1=1$ in Equation (2.1), where $\mu _{01}$ and $\mu _{02}$
are a complex conjugate pair of eigenvalues. Discretization is as in the
preceding example.

In this problem the starting direction of trajectories near $u_0$ is
unknown, making it difficult to generate a starting orbit in Step 0 with $%
u(1)$ in a small neighborhood of $u_1$. Indeed, this example also
illustrates the more global applicability of the successive continuation
approach.

\textbf{Step 0.} Initialize the problem parameters, $\gamma =.608$, $c=-.95$%
, and the algorithm parameters, $\epsilon _0=10^{-4}$, $\epsilon _1=.6283$, $%
T=10^{-2},$ and $c_1=1,$ $c_2=0.$ Initially $\func{Re}\mu _{01}=\func{Re}\mu
_{02}=1.508,$ $\func{Im}\mu _{01}=-\func{Im}\mu _{02}=1.388.$

\textbf{Step 1.}\QTR{textit}{\ } Compute a solution branch to (2.4), (2.7),
(2.11), (2.12a), (4.2), with scalar variables $T,\epsilon _1\in \Bbb{R}%
,\;d_0,d_1\in \Bbb{R}^3,$ $\tau _1,\tau _2\in \Bbb{R}$, in the direction of
increasing $T$ until $\tau _2$ crosses zero. There are ten free scalar
variables. Terminal values are $T=6.098,$ and $\epsilon _1=6.1738,$ $\tau
_1=-.9500.$

\textbf{Step 2.}\QTR{textit}{\ } Compute a solution branch to (2.4), (2.7),
(2.11), (2.12), (4.2), with scalar variables $T,\epsilon _1\in \Bbb{R}%
,\;d_0,d_1\in \Bbb{R}^3,$ $c_1,c_2,\tau _1\in \Bbb{R}$, in the direction of
increasing $T$. There are nine free scalar variables. Terminal values are $%
T=9.069,$ $\epsilon _1=1.737,$ and $\tau _1=-.3718.$

\textbf{Step 3}$.$ Fix $\epsilon _1$. Free $c\in \Bbb{R}$, and hence the
matrices $Q_0,Q_1\in \Bbb{R}^3\times \Bbb{R}^3$, and the entries $%
t_{11}^0,t_{12}^0,t_{13}^0,t_{21}^0,t_{22}^0,t_{23}^0,t_{33}^0$ and $%
t_{11}^1,t_{12}^1,t_{13}^1,t_{22}^1,t_{23}^1,t_{32}^1,t_{33}^1$ of the
matrices $T_0,T_1\in \Bbb{R}^3\times \Bbb{R}^3,$ respectively. Compute a
branch of solutions to the system (2.4), (2.5a), (2.6) - (2.10), and 
\begin{eqnarray}
t_1 &\equiv &t_{11}^0-t_{22}^0=0,  \tag{4.7a} \\
t_2 &\equiv &t_{22}^1-t_{33}^1=0,  \tag{4.7b}
\end{eqnarray}
in the direction of decreasing $\left| \tau _1\right| $ until $\tau _1$
crosses zero. This step involves 44 free scalar variables. Terminal values
are $c=-.8995,$ $T=7.436.$ Above, $t_{11}^0=\func{Re}\mu _{01},$ $t_{22}^0=%
\func{Re}\mu _{02},$ $t_{22}^1=\func{Re}\mu _{12},$ $t_{33}^1=\func{Re}\mu
_{13}.$

\textbf{Step 4}$.$ Fix $\tau _1=0$ and free $\epsilon _1\in \Bbb{R}$.
Compute a branch of solutions in the direction of decreasing $\epsilon _1,$
until $\epsilon _1=\epsilon _0$.The equations are as in Step 3. Terminal
values are $c=-.9027,$ $T=13.16,$ $\epsilon _1=10^{-4}.$

Note that, as in the general algorithm, it took three steps to get $\tau
_1=\tau _2=0$.

\subsection{Example 4: Heteroclinic orbits in a 4-D singular perturbation
problem.}

The existence of traveling wave front solutions to the singularly perturbed
reaction-diffusion system 
\begin{equation}
v_t=v_{xx}+v(v-a)(1-v)-w,\;\;\;\;w_t=\delta w_{xx}+\epsilon (v-\gamma w), 
\tag{4.8}
\end{equation}
for small positive $\epsilon $ and $\delta $, was established by Deng \cite
{d2}. In moving coordinates, $v_1=v(z),\;v_2=v^{^{\prime }}(z),$ $w_1=w(z),$ 
$w_2=w^{\prime }(z)$ with $z=t+cx,$ the reduced ODE is 
\begin{eqnarray}
v_1^{\prime } &=&v_2,  \tag{4.9a} \\
v_2^{\prime } &=&cv_2-v_1(1-v_1)(v_1-a)+w_1,  \tag{4.9b} \\
w_1^{\prime } &=&w_2,  \tag{4.9c} \\
w_2^{\prime } &=&[cw_2-\epsilon (v_1-\gamma w_1)]/\delta .  \tag{4.9d}
\end{eqnarray}

We compute a heteroclinic orbit with $u_0=(0,0,0,0)$ and $u_1=(\frac
12+\frac 12a+\frac 12\sqrt{(\gamma -2a\gamma +\gamma a^2-4)/\gamma },$ $0,$ $%
v_1/\gamma ,$ $0),$ with $\delta =\epsilon =.001$ and $a=.3$ fixed and and $%
\gamma $ and $c$ variable. Initially $\gamma =13.8$ and $c=.257$. In this
case $u_1=(.8736878,0,.0633107,0)$, and $n_0=n_1=2$, where the relevant
eigenvalues are $\mu _{0,1}=$.$6957$, $\mu _{0,2}=257.0537$, $\mu
_{1,1}=-.4415$, and $\mu _{1,2}=-0.0668$. Throughout we use a discretization
with 50 subintervals, 4 collocation points per interval and relative Newton
tolerances $10^{-8}$ for $u$ and $\lambda .$

In this problem there is strong divergence of trajectories near $u_0,$
making it difficult to generate a starting orbit in Step 0 with $u(1)$ in a
small neighborhood of $u_1$. Indeed, this example also illustrates the
continued applicability of the algorithm to such cases.

\textbf{Step 0.} Initialize the problem parameters, $\gamma =13.8$, $c=.2570$%
, and the algorithm parameters, $\epsilon _0=.6$, $\epsilon _1=.5161$, $%
T=10^{-5},$ $c_1=1,$ $c_2=0.$

\textbf{Step 1.}\QTR{textit}{\ } Compute a solution branch to (2.4), (2.7),
(2.11), (2.12a), (4.3), with scalar variables $T,$$\epsilon _1,\tau _1,\tau
_2\in \Bbb{R},\;d_0,d_1\in \Bbb{R}^4$, in the direction of increasing $T$
until $\tau _1$ crosses zero. Terminal values are $T=.0472$ and $\epsilon
_1=.5056$.

\textbf{Step 2.} Fix $\tau _1=0$, free $c_1$ and $c_2$, and compute a
solution branch to (2.4), (2.7), (2.11), (2.12), (4.3), in the direction of
increasing $T,$ with scalar variables $T,\epsilon _1,\tau _2,c_1,c_2\in \Bbb{%
R},\;d_1\in \Bbb{R}^4$. Final values are $T=1.363,$ $\epsilon
_1=.4024,\;c_1=1.00000,\;c_2=1.9\times 10^{-8},$ and $\tau _2=.521.$

\textbf{Step 3.}\QTR{textit}{\ }The scalar variables and equations are as in
Step 2, except that $T$ is fixed and $\epsilon _0$ is free. Compute a branch
of solutions in the direction of decreasing $\epsilon _0$ and $\epsilon _1$
to locate zero of $\tau _2$. Terminal values are $\epsilon
_0=.3199,\;\epsilon _1=.3991,\;c_1=1.00000,$and $c_2=1.3\times 10^{-8}$.
There is a very sensitive dependence on the ``shooting angle'', represented
by $c_2$.

\textbf{Step 4.}\QTR{textit}{\ }The free parameters\QTR{textit}{\ }$%
c_1,c_2\in \Bbb{R}$ are replaced by $d_0\in \Bbb{R}^4.$ Fix $\tau _2=0$ and
free $T$. Also free the problem parameter $c\in \Bbb{R}$, and hence the
matrices $Q_0,Q_1,T_0,T_1\in \Bbb{R}^4\times \Bbb{R}^4.$ Compute a solution
branch to the system (2.4)-(2.10) in the direction of increasing $T$.
Terminal values are $c=.3437209,$ $T=12.66,$ and $\epsilon _0=10^{-4}.$

\textbf{Step }$\mathbf{5.}$ The scalar variables and equations are as in
Step 4, except that $\epsilon _0$ is fixed and $\epsilon _1$ is free.
Compute a solution branch in the direction of increasing $T$. Final values
are $c=.2572501,$ $T=71.31,\;\epsilon _1=5\times 10^{-3}$. \newline

\section{Discussion}

The initial detection of a connecting orbit in a dynamical system is
generally a difficult task; often with extremely sensitive dependence on
initial conditions and on parameter values. This is even more so in the case
of singularly perturbed equations. Thus there is use for a systematic
procedure that has a good chance of success, even in difficult problems. In
this paper we have presented a successive continuation method for locating
connecting orbits. We have shown that the procedure works, provided the
connecting orbit is isolated and the initial orbit sufficiently close. Thus
our analysis only guarantees local convergence of the method, even though it
has been designed for extended convergence properties, as is well
illustrated by our numerical examples, which include some ``hard'' problems.
A complete presentation of global convergence properties is beyond the scope
of this paper, and perhaps best presented in a more general context. The key
ideas in a global analysis of the successive continuation algorithm are the
following. It is assumed that the problem can be reduced to that of finding
a small number of parameters for which an equal number of equations is to be
satisfied. Given a subset of the equations that are satisfied, a
one-dimensional continuum of such solutions is followed, and points where
any of the remaining equations are satisfied are accurately located.
Proceeding, inductively, one then continues an enlarged set of satisfied
equations. At any stage of the algorithm, the continuation procedure works
locally, provided the solution points are ``regular'' \cite{dkk}.
Generically this is the case, while for nongeneric problems (which are often
encountered!) one can regularize the problem by adding (if necessary)
unfolding parameters and (again, if necessary) regularizing equations. In
fact, the method of pseudo-arclength continuation itself is the simplest
nontrivial example of this procedure.

The power of the successive continuation procedure is then its ability to
reach a solution from a far away starting point, and, in fact, to locate
multiple solutions and often all solutions, provided the regularity
assumptions are satisfied and provided \textit{the solution(s) are reachable}%
. It is the latter condition that need not always be satisfied. In fact, it
is easy to construct simple (algebraic) examples where a solution is not
reachable from a given starting point. One could argue that this problem can
also be solved by adding unfolding parameters, but in practice it is not
often clear how to do this, for example, in the case of computing connecting
orbits. Nevertheless, as our numerical experience has shown, including the
examples presented in this paper, the successive continuation method
provides, at least, a useful tool that often gives results where other
methods fail.

Note that, in our computation of connecting orbits, the above reduction to a
``small-dimensional problem'' does not require the problem to be posed as a
``shooting problem'', which would make the algorithm useless, for example,
for the singular perturbed equation in Section~4.1. Throughout, the small
dimensional problem remains embedded in the full set of equations, which are
solved by continuation in the full space. Even the initial ``integration''
(see Algorithm, Step 1) is done by continuation of complete orbits, in order
to be able to monitor, and react to (e.g. adaptive mesh refinement),
sensitive dependence, for example, on the left boundary conditions.

Although, as pointed out above, a discussion of global convergence
properties is perhaps best presented in terms of the global topology of the
underlying manifolds, one can also obtain global results by making certain
global assumptions on the vector field, although this is practically less
useful in the present context. In fact, applying the theory in \cite{k}, one
can show the following. Let $F_h^{(n_\tau )}:X_h\rightarrow Y_h\times \Bbb{R}%
^{n_\tau }$ be the discretization \cite[pp. 748-749]{dkk} of $F^{(n_\tau
)}:X\rightarrow Y\times \Bbb{R}^{n_\tau }$ defined by (3.11b). Let $\Omega
\subset X_h$ be a bounded open set with a smooth connected boundary $%
\partial \Omega .$ Suppose that the operator $F_h^{(n_\tau )}$ satisfies
Smale boundary conditions \cite{k}

a) $D_UF_h^{(n_\tau )}(U_h,\Lambda _h)$ is nonsingular on $\partial \Omega ;$
and either

b) $(D_UF_h^{(n_\tau )}(U_h,\Lambda _h))^{-1}F_h^{(n_\tau )}(U_h,\Lambda _h)$
points into $\Omega ,$ $\forall U_h\in \partial \Omega ;$ or

c) $(D_UF_h^{(n_\tau )}(U_h,\Lambda _h))^{-1}F_h^{(n_\tau )}(U_h,\Lambda _h)$
points out of $\Omega ,$ $\forall U_h\in \partial \Omega .$

Then by a slight generalization of \cite[Th. 2.4]{k} one can show that $%
\forall U_h^0\in \partial \Omega $ there exists a piecewise smooth path with
starting point $(U_h^0,\Lambda _h^0)$ and terminal point $(U_h^{*},\Lambda
_h^{*})$ with $F_h^{(n_\tau )}(U_h^{*},\Lambda _h^{*})=0.$

\textbf{Acknowledgment.} The authors wish to thank Professor J. Demmel,
University of California, Berkeley, for helpful discussions concerning the
real Schur factorization, and Dr. A.~Champneys, University of Bristol, for
helpful comments on an earlier version of this paper.


\begin{thebibliography}{99}
\bibitem{ahls}  N. Aubry, P. Holmes, J.L. Lumley, and E. Stone, The dynamics
of coherent structures in the wall region of a turbulent boundary layer, 
\textit{J. Fluid Mechanics}, \textbf{192} (1988), 115-173.

\bibitem{bn}  H. Beresticky and L. Nirenberg, Traveling fronts in cylinders, 
\textit{Ann. Inst. Henri Poincar\'{e}}, \textbf{9}, No. 5 (1992), 497-572.

\bibitem{b1}  W.-J. Beyn, The numerical computation of connecting orbits in
dynamical systems, \textit{IMA J. Numer. Anal.}\QTR{textit}{, }\textbf{9}
(1990), 379-405.

\bibitem{b2}  W. -J. Beyn, Global bifurcations and their numerical
computation, in: D. Roose\textit{\ et al}\QTR{textit}{.}, Ed., \textit{%
Continuation and Bifurcations: Numerical Techniques and Applications}%
\QTR{textit}{, }Kluwer, Dordrecht, The Netherlands (1990), 169-181.

\bibitem{bfmp}  D.L. Brown, M.G. Forest, B.J. Miller and N.A. Petersson,
Computation and stability of fluxons in a singularly perturbed sine-Gordon
equation, \textit{SIAM J. Appl. Math}\QTR{textit}{.}\QTR{textbf}{\ }\textbf{%
54}, No. 4 (1994), 1048-1066.

\bibitem{c}  V. Canale, The computation of paths of homoclinic orbits, 
\textit{Ph.D. Thesis}\QTR{textit}{,} Department of Computer Science,
University of Toronto, Toronto, Canada (1994).

\bibitem{ck}  A.R. Champneys and Yu.A. Kuznetsov, Numerical detection and
continuation of codimension-two homoclinic orbits,\textit{\ Int. J. Bif. and
Chaos}\QTR{textit}{:}\textbf{\ 4}, No. 4 (1994), 785-822.

\bibitem{cks}  A.R. Champneys, Yu.A. Kuznetsov and B. Sandstede, A numerical
toolbox for homoclinic bifurcation analysis, \textit{Applied Nonlinear
Mathematics Research Report}\QTR{textit}{\ No. 1.95, University of Bristol}
(1995), \QTR{textit}{Bristol}, UK.

\bibitem{cwhz}  L. O. Chua, C. W. Wu, A. Huang, and C. Q. Zhong, A universal
circuit for studying and generating chaos, Part I : Routes to chaos, \textit{%
IEEE Trans. on Circuits and Systems-I : Fundamental Theory and Applications }%
\textbf{10} (40), (1993), 732-744.\QTR{textit}{\ \textit{Special Issue on
Chaos in Electronic Circuits}, Part A.}

\bibitem{d}  J. Dieudonn\'{e},\textit{\ Foundations of Modern Analysis}%
\QTR{textit}{,} Academic Press (1960).

\bibitem{d1}  B. Deng, Constructing Homoclinic Orbits and Chaotic Attractors,%
\textit{\ Int. J. Bifurcation and Chaos},\textbf{\ 4}, No. 4, (1994),
823-841.

\bibitem{d2}  B. Deng, personal communication (1995).

\bibitem{df}  E.J. Doedel and M.J. Friedman, Numerical computation of
heteroclinic orbits, \textit{J. Comput. and Appl. Math.}\QTR{textit}{,} 
\textbf{26 }(1989), 159-170.

\bibitem{dfg}  E.J. Doedel, M.J. Friedman, and J. Guckenheimer, On computing
connecting orbits: general algorithm and application to the Sine-Gordon and
Hodgkin-Huxley equations, \textit{The IEICE Transactions on Fundamentals of
Electronics, Communications and Computer Science} \textbf{77}, No. 11
(1994), \textit{Special Section on Nonlinear Theory and Its Applications}.

\bibitem{dfm}  E.J. Doedel, M.J. Friedman, and A.C. Monteiro. On locating
connecting orbits, \textit{Appl. Math. and Comput. }\textbf{65}, No. 1-3
(1994), 231-239.

\bibitem{dkk}  E.J. Doedel, H.B. Keller and J.P. Kern\'{e}vez, Numerical
Analysis and Control of Bifurcation Problems, Part I: Bifurcation in Finite
Dimensions, \textit{Int. J. Bif. and Chaos}\QTR{textit}{:} \textbf{1}, No. 3
(1991), 493-520; \QTR{textit}{\ Part II:} Bifurcation in Infinite Dimensions,%
\textit{\ Int. J. Bif. and Chaos}\QTR{textit}{:} \textbf{1}, No. 4 (1991),
745-772.

\bibitem{dwf}  E.J. Doedel, X.J. Wang and T.F. Fairgrieve, \textit{AUTO 94:
Software for continuation and bifurcation problems in ordinary differential
equations,} Applied Mathematics Report, California Institute of Technology
(1994).

\bibitem{fra}  E. Freire, L.G. Franquelo and J. Aracil, Periodicity and
chaos in an autonomous electronic circuit,\textbf{\ }\textit{IEEE
Transactions on Circuits and Systems}\textbf{, }CAS \textbf{31}, No. 3
(1984), 237-247.

\bibitem{frgp}  E. Freire, A.J. Rodr\'{\i }guez-Luis, E. Gamero and E.
Ponce, A case study for homoclinic chaos in an autonomous electronic
circuit, \textit{Physica D} \textbf{62} (1993), 230-243.

\bibitem{f}  M.J. Friedman, Numerical analysis and accurate computation of
heteroclinic orbits in the case of center manifolds,\QTR{textit}{\ \textit{%
J. Dynam. Diff. Eq}.} \textbf{5}, No. 1 (1993), 59-87.

\bibitem{fd1}  M.J. Friedman and E.J. Doedel, Numerical computation and
continuation of invariant manifolds connecting fixed points,\QTR{textit}{\ 
\textit{SIAM, J. Numer. Anal}.}\QTR{textbf}{\ }\textbf{28,} No. 3 (1991),
789-808.

\bibitem{fd2}  M.J. Friedman and E.J. Doedel, Computational methods for
global analysis of homoclinic and heteroclinic orbits: a case study, \textit{%
J. Dynam. Diff. Eq}\QTR{textit}{.}\textbf{\ 5}, No. 1 (1993), 37-58.

\bibitem{gv}  G.H. Golub and C.F. Van Loan, \textit{Matrix Computations},
The John Hopkins University Press (1989).

\bibitem{gs}  P. Gray and S.K. Scott, \textit{Chemical Oscillations and
Instabilities: Nonlinear Chemical Kinetics}, Oxford, (1990).

\bibitem{k}  H.B. Keller, Global homotopies and Newton's methods, in 
\QTR{textit}{Recent Advances in Numerical Analysis}, Academic Press (1978),
73-94.

\bibitem{l}  S. Lang, \textit{Introduction to Differential Manifolds},
Interscience Publications (1962).

\bibitem{lk}  M. Lentini and H.B. Keller, Boundary value problems over semi
infinite intervals and their numerical solution, \textit{SIAM J. Numer. Anal}%
\QTR{textit}{. }\textbf{17} (1980), 557-604.

\bibitem{mt}  B.A. Malomed and R.S. Tasgal, Vibration modes of a gap soliton
in nonlinear optical medium, \textit{Phys. Rev. E} \textbf{49}, No. 6
(1994), 5787-5796.

\bibitem{m}  F.C. Moon, \QTR{textit}{Chaotic Vibrations: }\textit{An
Introduction for Applied Scientists and Engineers}, Wiley, N.Y. (1987).

\bibitem{re}  J. Rinzel and G.B. Ermentrout, Analysis of neural excitability
and oscillations. In C. Koch and I. Sedev (eds.), \textit{Methods in Neural
Modeling: From Synapses to Networks.} MIT Press, Cambridge, MA, (1989).

\bibitem{rlfp}  A.J. Rodr\'{\i }guez-Luis, E. Freire and E. Ponce, A method
of homoclinic and heteroclinic continuation in two and three dimensions, in:
D. Roose \QTR{textit}{et al.}, eds.,\textit{\ Continuation and Bifurcations:
Numerical Techniques and Applications,}\QTR{textit}{\ }Kluwer, Dordrecht,
The Netherlands (1990), 197-210.

\bibitem{sa}  B. Sandstede, Convergence estimates for the nunerical
approximation of homoclinic solutions, \textit{Preprint} No.~198,
Weierstrass-Institut f\"{u}r Angewandte Analysis und Stochastik, Berlin,
1995.

\bibitem{s}  S. Schecter, Numerical computation of saddle-node homoclinic
bifurcation points, \textit{SIAM J. Numer. Anal}\QTR{textit}{. }\textbf{30},
No. 4 (1993), 1155-1178.
\end{thebibliography}
\end{document}